\input{aipcheck}

\documentclass[
    ,final            
  ]
  {aipproc}

\layoutstyle{6x9}

\usepackage{bm}
\newcommand{\vecc}[1]{\mbox{\boldmath $#1$}}

\begin{document}

\title{Wilson lines in the operator definition of TMDs: \\
       spin degrees of freedom and renormalization\footnote{Talk given at the XIX International Workshop on Deep Inelastic Scattering (DIS 2011),
11 - 15 Apr 2011, Thomas Jefferson National Accelerator Facility, Newport News (VA)}}


\author{I.O. Cherednikov}{
  address={Departement Fysica, Universiteit Antwerpen,
         B-2020 Antwerpen, Belgium}, altaddress={Joint Institute for Nuclear Research, RU-141980 Dubna, Russia}
}

\author{A.I. Karanikas}{
  address={Department of Physics, University of Athens,
           Nuclear and Particle Physics Section,
           Panepistimiopolis, GR-15771 Athens, Greece}
}

\author{N.G. Stefanis}{
  address={Institut f\"{u}r Theoretische Physik II,
           Ruhr-Universit\"{a}t Bochum,
           D-44780 Bochum, Germany}
} 

\begin{abstract}
A generalized idea of gauge invariance, that embodies into the Wilson
lines the spin-dependent Pauli term
$\sim F^{\mu\nu}[\gamma_\mu, \gamma_\nu]$, is applied to set up a new
framework for the operator definition of transverse-momentum-dependent
parton densities (TMDs).
We show that such a treatment of gauge invariance is justified,
since it does not change the leading-twist behavior of the TMDs,
albeit it contributes to their twist-three properties, in particular,
to their anomalous dimensions.
We discuss other consequences of this generalization and its possible
applications to lattice simulations of the TMDs.
\end{abstract}

\hfill {\small{\tt RUB-TPII-05/2011}}

\maketitle

In order to render operator products and their (hadronic) matrix
elements gauge-invariant, one usually uses a path- ($\mathcal{C}$)
dependent gauge link (Wilson line) with an exponent
containing only the gauge field $A$:
\begin{equation}
  [y;x|\mathcal{C}]
\equiv
  \mathcal{P} \exp
                  \left[
                  - ig \int_{x[\mathcal{C}]}^{y}
                           dz^\mu A_{\mu}^{a}(z)t^a
                  \right] \ .
\label{eq:gaugelink}
\end{equation}
This is, however, the {\it minimal} option which simply reflects
the fact that color vectors cannot be compared at a distance.
The gauge potential $A_{a}^{\mu}$ as such is spin-blind; hence one
loses any information about the transfer of the spin degrees of
freedom along the contour $\mathcal{C}$.
Therefore, to include a direct spin interaction, one has to include
into the gauge link an additional term proportional to the gluon
strength tensor $F_{\mu\nu}^{a}$ (the so-called Pauli term) that
explicitly accommodates the spin-dependent interactions.
These non-minimal, i.e., {\it enhanced} gauge links, generalized with
the inclusion of the Pauli contribution
$\sim F_{\mu\nu}^{a} J_{\mu\nu}$
(with $J_{\mu\nu}=(1/4)[\gamma_\mu, \gamma_\nu]$)
are normally ignored.
This simplification appears natural in the case of, e.g., integrated
(collinear) parton distribution functions, where the integration path is
trivial and goes along a straight lightlike line.

On the other hand, operator definitions of the unintegrated
transverse-momentum dependent parton distributions (TMDs)
\cite{TMD_basic} contain a compound of longitudinal and
transverse (at light-cone infinity) gauge links, the color
structure and the space-time setup of which may be rather
complicated \cite{TMD_gauge_links}.
In the latter case, the non-triviality of the integration
contours makes it crucial to take into account contributions
of the non-minimal spin-dependent terms.
In lattice realizations of TMDs, the spin-dependent terms in the
Wilson lines can also produce a significant effect for various choices
of the integration path on the lattice \cite{TMD_Lattice, TMD_INT}.

To justify the introduction of the enhanced Wilson lines, let us
imagine two orthogonal ``spaces'', with a cross-talking between a
pair of quantum fields.
The first ``space'' is the color space, where such a binary relation
is accomplished in terms of the {\it minimal} Wilson lines in the
fundamental or adjoint representation of $SU(3)_{\rm c}$.
The spin correlations (generated by the Pauli terms) are, in contrast,
defined in the second ``space''.
From straightforward power-counting we conclude that the spin-dependent
terms are of the nonleading-twist with respect to the spin-blind ones.
However, our analysis demonstrates that the inclusion of the Pauli
terms, though invisible in the completely unpolarized TMDs, can affect
significantly a number of polarized distributions, e.g., those
responsible for time-reversal-odd phenomena, such as single-spin
asymmetries \cite{TMD_INT}.
Adopting this encompassing idea of gauge invariance, we have to clarify
whether the definition of TMDs, we proposed before in
Refs.\ \cite{CS_all}, has to be modified and to which extent the
incorporation of the Pauli term has phenomenological consequences, for
instance, for the UV evolution of TMDs.

In this talk, we report on the first results of our recent study of
the renormalization-group properties (anomalous dimensions) of the
TMD distribution functions with enhanced gauge links \cite{CKS10}.
Because the UV properties of this matrix element are independent of
specific hadronic states, we consider the ``quark-in-a-quark'' TMD.
According to our generalized concept of gauge invariance, the
{\it unsubtracted} distribution function (i.e., that without a soft-term
supplement) of a quark with momentum $k$ and flavor $i$ in a quark with
momentum $p$ reads
\begin{eqnarray}
& &
{\cal F}_{i/q}^{\Gamma}(x, \mathbf{k}_{\perp})
 = {}
   \frac{1}{2} {\rm Tr} \! \int\! dk^-
   \int \! \frac{d^4 \xi }{(2\pi)^4}\,
   {\rm e}^{- i k \cdot \xi}
\langle  p \ |\bar \psi_i (\xi) [[\xi^-, \bm{\xi}_{\perp};
   \infty^-, \bm{\xi}_{\perp}]]^\dagger \nonumber \\
& &
 \times
 [[\infty^-, \bm{\xi}_{\perp};
   \infty^-, \bm{\infty}_{\perp}]]^\dagger  \Gamma
\left.
   [[\infty^-, \bm{\infty}_{\perp};
   \infty^-, \mathbf{0}_{\perp}]]
   [[\infty^-, \mathbf{0}_{\perp};
   0^-, \mathbf{0}_{\perp}]]
\psi_i (0) | p \right \rangle
\vspace{-1cm}
\label{eq:TMD-PDF}
\end{eqnarray}
where $\Gamma$ denotes one or more $\gamma$-matrices and corresponds
to the particular distribution under consideration.
The state $|p\rangle$ stands for the quark target state.

An important comment about definition (\ref{eq:TMD-PDF}) is in order.
We started from the ``fully unintegrated'' correlation function, which
depends on \textit{all} four components of the parton's momentum
\cite{UNint_full}.
Thus, the TMD PDF is obtained \textit{after} performing the $k^-$
integration that formally renders the coordinate $\xi^+$ equal to
zero
$
  \int\! dk^- {\rm e}^{- i k^- \xi^+}
=
  2\pi \delta (\xi^+)
$.
However, this operation may produce additional divergences because,
carrying it out, all quantum fields involved (quarks and gluons) are
defined on the light ray $\xi^+ = 0$.
This means that the plus light-cone coordinates of the product of two
quantum fields always coincide.
The appropriate treatment of this delicate issue is discussed in detail
in \cite{CKS10}.

We define the enhanced {\it longitudinal} gauge link along the
$x^-$ direction:
\begin{equation}
  [[\infty^-, \mathbf{0}_{\perp}; 0^-, \mathbf{0}_{\perp}]]
 =
  \mathcal{P}
  \exp
      \left[
            - ig \int_{0}^{\infty} d\sigma \ u_{\mu} \
                 A_{a}^{\mu}(u \sigma)t^a
           - i g \int_{0}^{\infty} d\sigma  \
                 J_{\mu\nu} F_{a}^{\mu\nu}(u \sigma)t^a
      \right]
\label{eq:lightlike-link}
\end{equation}
and the enhanced {\it transverse} gauge link:
\begin{equation}
  [[\infty^-,  \bm{\infty}_{\perp}; \infty^-, \mathbf{0}_{\perp}]]
 =
  \mathcal{P}
  \exp
      \left[
            - ig \int_{0}^{\infty} d\tau
            \mathbf{l}_{\perp} \! \cdot \!
            \mathbf{A}_{\perp}^{a}(\vecc l\tau)t^a
            - i g \int_{0}^{\infty} d\tau
            J_{\mu\nu}F_{a}^{\mu\nu}(\vecc l\tau)t^a
      \right] \ ,
\label{eq:transverse-link}
\end{equation}
where the two-dimensional vector
$\mbox{\boldmath $l$}\equiv \vecc l_\perp$
is arbitrary.
We make use of the following reparameterization of the
(initially dimensionless) vectors defining the path of the integration
as
$
  n^+_\mu \to u^*_\mu
=
  \frac{1}{p^+} n^+_\mu  \ \ , \ \ n^-_\mu \to u_\mu =  p^+ n^-_\mu \
$,
which implements boosts in the longitudinal directions.
The plus-component of the momentum $p$ is large in the given
kinematics and is the only momentum scale in our
reparameterization\footnote{An analogous definition holds for the
$x^+$ direction, provided one makes the replacement $u \to u^*$.}.
The enhanced Wilson lines, introduced above, viz.,
Eqs.\ (\ref{eq:lightlike-link}) and (\ref{eq:transverse-link})
set up a corner stone of the concept of generalized gauge invariance
in the operator formalism of the TMDs.

\begin{figure*}[ht]
  \includegraphics[width=0.45\textwidth,height=0.65\textheight,angle=90]{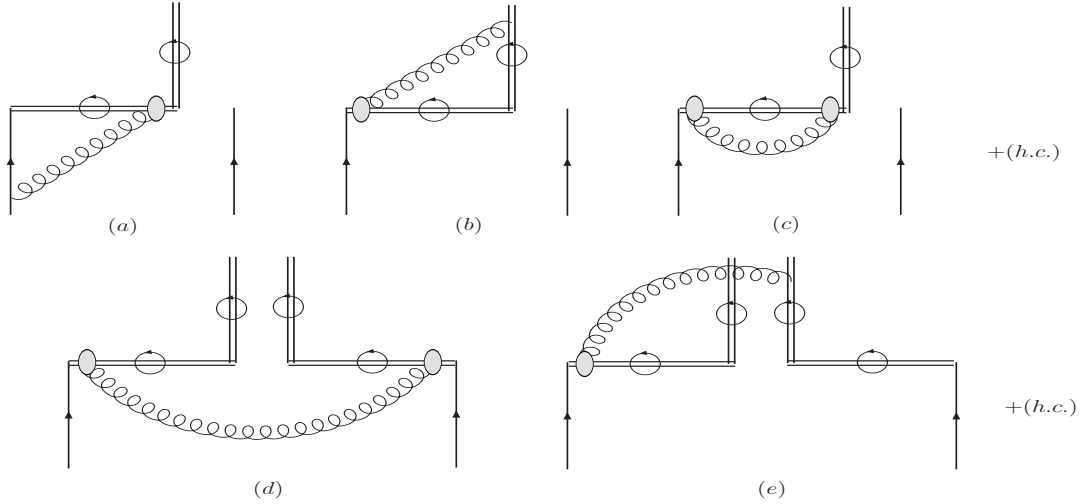}
  \caption{One-loop Feynman graphs contributing to the TMD (\ref{eq:TMD-PDF}).
Double lines denote minimal gauge links, while those with a ring
represent enhanced gauge links with Pauli contributions.
Fermions and gluons are shown as solid and curly lines, respectively.
Graphs (a), (b), (c), and (d) describe virtual gluon corrections;
graphs (e), (f), and (g) represent real-gluon exchanges.
\protect\label{fig:graphs}}
\end{figure*}

\begin{figure*}[ht]
  \includegraphics[width=0.5\textwidth,height=0.67\textheight,angle=90]{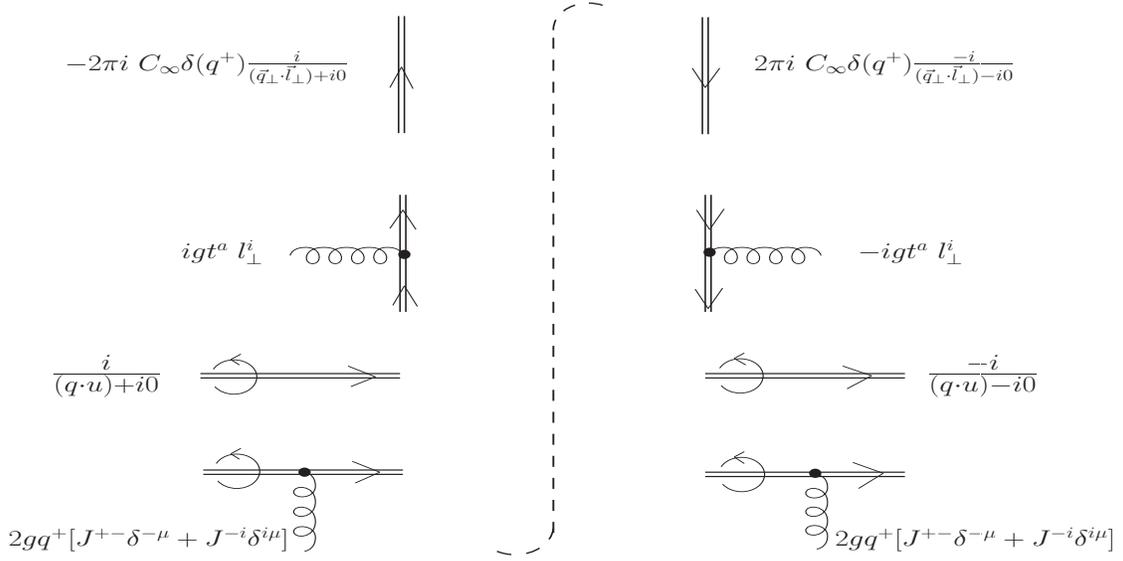}
  \caption{Feynman rules for the calculation of the one-loop graphs
  shown in Fig.\ \ref{fig:graphs} in the light-cone gauge with enhanced
  gauge links.
Rules are given for both sides of the final-state cut (long vertical
line).
Vertical double lines denote the transverse gauge links at
light-cone infinity; the horizontal ones with arrowed rings are
the spin-dependent lightlike enhanced gauge links.
\protect\label{fig:pauli-feynman}}
\end{figure*}

Let us now briefly describe some of the quantitative results obtained
within this framework (see \cite{CKS10} and our previous works
in \cite{CS_all} for technical details).
To analyze the UV divergences in the leading $\alpha_s$-order, one has
to evaluate the diagrams displayed in Fig.\ \ref{fig:graphs}.
The corresponding Feynman rules are summarized in
Fig.\ \ref{fig:pauli-feynman}.

As a result, the non-trivial UV-singular (in the dimensional
regularization) contribution arises from the cross-talking of the gauge
fields belonging to the transverse {\it minimal} and longitudinal
{\it enhanced} Wilson lines, e.g., from diagram Fig. 1(d).
In the case of the twist-two Dirac structures
$
 \Gamma_{\rm tw\,-\,2}
=
 \{ \gamma^+, \gamma^+\gamma^5, i\sigma^{i+}\gamma^5 \}
$,
the corresponding singular terms cancel by their ``mirror''
(Hermitean conjugated) counterparts.
In contrast, the twist-three TMDs
(say, $\Gamma_{\rm tw\,-\,3} = \gamma^i$)
get from these terms non-trivial UV divergent contributions of the type
\begin{equation}
  \Gamma_{\rm tw\,-\,3} \langle \mathbf{A}^{\perp} F^- \rangle
  + \langle \mathbf{A}^{\perp} F^- \rangle \Gamma_{\rm tw\,-\,3}
=
  - C_{\rm F} \ \frac{1}{4\pi} [\gamma^+, \gamma^-]
  \ \Gamma (\varepsilon) \ \left( 4\pi \frac{\mu^2}{\lambda^2}\right)^\varepsilon \ ,
\label{eq:uv_singular}
\end{equation}
where $\langle \mathbf{A}^{\perp} F^- \rangle$ stands for the result of the
calculation of the diagram in Fig. 1(d).

To conclude, we presented a new framework \cite{CKS10} for completely
gauge-invariant TMDs which takes into account explicitly the spin
degrees of freedom of quantum particles by means of the Pauli term in
the Wilson lines.
Let us stress that because the spin-dependent terms contribute to the
UV anomalous dimensions of the twist-three TMDs, their evolution
appears to be non-trivial, compared to the RG-properties of the TMDs
with minimal Wilson lines.
This calls for the modification of the renormalization procedure to
preserve the parton-number interpretation of TMDs that deserves a
dedicated investigation in the future.


\bibliographystyle{aipproc}   

\end{document}